\documentclass[twocolumn,showpacs,preprintnumbers,amsmath,amssymb]{revtex4}
\usepackage[dvips]{graphicx}
\usepackage{graphics}
\usepackage{epsf}
\begin{document}

\newcommand{\e}[1]{\times 10^{#1}}

\newcommand{\fig}[2]{
\begin{figure}[htp]
\centerline{\epsfxsize 0.99\columnwidth \epsfbox{#1}}
\caption{#2
}
\end{figure}
}
\newcommand{\figtwo}[3]{
\begin{figure}[htp]
\centerline{\epsfxsize 0.99\columnwidth \epsfbox{#1}}
\centerline{\epsfxsize 0.99\columnwidth \epsfbox{#2}}
\caption{#3
}
\end{figure}
}

\title{Kinetic Monte Carlo simulation of faceted islands in
  heteroepitaxy using multi-state lattice model}

\author{Chi-Hang Lam}
\affiliation{
Department of Applied Physics, Hong Kong Polytechnic University,
Hung Hom, Hong Kong, China}

\date{\today}

\begin{abstract}
  A solid-on-solid model is generalized to study the formation of Ge pyramid islands bounded by (105) facets on Si(100) substrates in two dimensions. Each atomic column is not only characterized by the local surface height but also by two deformation state variables dictating the local surface tilt and vertical extension. These deformations phenomenologically model surface reconstructions in (105) facets and enable the formation of islands which better resemble faceted pyramids. We demonstrate the model by application to a kinetic limited growth regime. We observe significantly reduced growth rates after faceting and a continuous nucleation of new islands until overcrowding occurs.
\end{abstract}

\pacs{68.65.Hb, 81.16.Dn, 81.16.Rf}
\maketitle

\section{Introduction}

Strain induced self-assembly of three dimensional (3D) islands in
heteroepitaxy have been attracting much research interest because of
the rich physics involved and their potential applications as quantum
dots in optoelectronic devices
\cite{Politi2000,Shchukin2003,Berbezier2009}. A widely studied system
is Ge deposited on Si(100) substrates with a 4\% lattice
misfit. Relatively flat islands in the form of stepped mounds with
unfaceted sidewalls called pre-pyramids start to emerge at 3
monolayers (MLs) of Ge coverage \cite{Mo1990,Vailionis2000}. Further
deposition leads to pyramids or rectangular-based huts bounded
by (105) facet planes. Deposition temperatures lower than 500$^\circ
C$ generally favors rectangular huts \cite{Kastner1999,Drucker2008}
while higher temperature often leads to pyramids \cite{Rastelli2003}.
After still further deposition or annealing, pyramids can grow into
dome islands bounded mainly by steeper (113) facets
\cite{Medeiros1998,Rastelli2005}.

(105) facets on pyramids and huts have been found to be
extraordinarily stable and atomically flat from first principle
calculations
\cite{Fujikawa2002,Montalenti2005,Lu2005,Shklyaev2005}. At low
temperature, surface steps on (105) facets are rarely observed
\cite{Drucker2008}. They are however present at higher temperature and
the bunching of them are observed to be important to the morphological
evolution \cite{Montalenti2004}. The structures, energies and dynamics of
these steps have been studied using first-principles calculations
\cite{Montalenti2007}. Also, the edge energies of a (105) faceted
ridge have been estimated using molecular dynamics simulations based on
empirical potentials \cite{Retford2007}.

Large scale simulations of the formation of 3D islands is possible
using kinetic Monte Carlo (KMC) methods based on lattice models
\cite{Orr1992,Khor2000,Meixner2001,Lam2002,Gray2005,Lung2005,
  Smereka2006,Lam2008,Smereka2009,Zhu2007}. The simulations are
computationally very intensive due to the long-range nature of elastic
interactions.  Elastic forces can be accounted for accurately and
efficiently using advanced algorithms so that simulations in 2D
\cite{Lam2002,Gray2005} and 3D
\cite{Lung2005,Smereka2006,Lam2008,Smereka2009} with respectively
large and moderate system sizes are possible. Using more approximate
forms of the elastic interactions, larger systems in 3D can also be
studied \cite{Meixner2001,Zhu2007}.

KMC studies on strained layers are generally based on square or cubic
lattices for simplicity. Strain induced islands or pits are readily
generated but their sidewalls are almost vertical
\cite{Orr1992,Smereka2006,Smereka2009} or at an of inclination of
about 45 degrees \cite{Khor2000,Lam2002,Lung2005,Zhu2007} depending on
the details of the bond energies or additional constraints used. These
inclinations are much steeper than 11$^\circ$ and 26$^\circ$ for the
(105) and (113) facets respectively. The realistic facets however are
of rather low-symmetry and in general are not favored energetically in
lattice models.  The discrepancy results in strain distributions
considerably different from the realistic ones and may probably lead
to qualitatively different growth modes in certain
situations. Furthermore, the surface energy of the island sidewalls
from existing KMC models are not independently adjustable and there is
no simple approach to incorporate for example the extraordinary
stability of certain facets.  In addition, with only one favored sidewall
slope in a given model, only one type of island can be simulated so
that studying the pyramid to dome transition for instance is
impossible.

In this work, we extend the convectional ball and spring lattice
model for KMC simulation of heteroepitaxial solids in 2D by allowing
specific geometrical deformation states of the surface atoms. These
deformations phenomenologically represent surface reconstructions on
(105) facets. We show computationally that this new multi-state model
leads to the formation of faceted islands. Examples of qualitative
differences in the growth dynamics between faceted and unfaceted
islands are explained.

\section{Ball and spring Lattice Model}
\label{S:MBE}

We first explain the conventional square lattice model of elastic
solids in 2D while further extensions will be introduced in the next
section. Every atom is associated with a lattice sites and are
connected to nearest and next nearest neighbors by elastic
springs. Solid-on-solid conditions are assumed. We follow the model
parameters used in Ref. \cite{Lam2002} unless otherwise stated to
approximate the widely studied Ge/Si(001) system.  We assume a
substrate lattice constant $a_s=2.715$\AA ~ so that $a_s^3$ gives the
correct atomic volume in crystalline silicon. The lattice misfit
$\epsilon=(a_f-a_s)/a_f$ equals 4\% where $a_f$ is the lattice
constant of the film.
Nearest and next nearest neighboring atoms are directly connected by
elastic springs with force constants $k_N=13.85eV/a_s^2$ and
$k_{NN}=k_N/2$ respectively. 
The elastic couplings of adatoms with the rest of the system
are weak and are completely neglected for better computational
efficiency.  In this model, surface steps have
a particularly high tendency to bunch together under strain presumably
due to the much weaker entropic surface step repulsion in 2D. We hence
forbid double surface steps as well as adjacent single surface steps
of the same direction so that the steepest surface slope allowed is 1/2.

The KMC approach simulates the morphological evolution by explicitly
considering the diffusion of surface atoms. Every topmost atom $m$ on
the film can hop to a nearby site with a hopping rate $\Gamma(m)$
following an Arrhenius form:
\begin{equation}
\label{rate}
\Gamma(m) = 
{R_0}\exp \left[ -\frac{n_m \gamma  - \Delta E_s(m) - E_0}{k_{B}T}\right]
\end{equation}
where $n_m$ is the number of nearest and next nearest neighbors of
atom $m$. We have assumed an identical nearest and next nearest
neighbor bond strength $\gamma$. We put $\gamma=0.5 eV$, slightly larger
than the value in Ref. \cite{Lam2002} so that the energy costs of stepped
mounds become slightly higher.
The energy $\Delta E_s(m)$ is the difference in the strain
energy $E_s$ of the whole lattice at mechanical equilibrium with or
without the atom $m$. Due to the long-range nature of elastic
interactions, its efficient calculation is highly nontrivial and we
handle it using a Green's function method together with a
super-particle approach explained in Refs. \cite{Lam2002,Lam2008,Lung2007}.
In addition, $E_0=3\gamma-0.67eV$, where 0.67eV is the adatom
diffusion barrier on the (100) plane.  To speed up the simulations,
long jumps are allowed so that a hopping atom will jump directly to
another random topmost site at most $s_{max}=8$ columns away with equal
probability. Then, $R_0=2D_0/(\sigma_s a_s)^2$
with $D_0=3.83\times 10^{13}\mbox{\AA}^2 s^{-1}$ and $\sigma_s^2 =
\frac{1}{6}(s_{max}+1)(2s_{max}+1)$.  This gives the appropriate
adatom diffusion coefficient for silicon (100).

\section{Multi-state Lattice model with Surface deformation}
\label{S:extended}

To effectively model (105) facets, which are more precisely (15)
surfaces in 2D, we introduce additional degrees of freedom
representing local deformations to all topmost atoms. They
phenomenologically accounts for the surface rebonding or
reconstruction states on a (105) faceted region \cite{Fujikawa2002}.
For efficient computation, these deformations localized to individual
surface atoms are assumed to be completely independent of the lattice
misfit, although correlations between misfit strain and surface
reconstruction are known to exist,
\cite{Montalenti2005,Lu2005,Shklyaev2005}. In the following
calculation of the local deformation energies, we hence neglect
lattice misfit and express all lengths in unit of lattice constant. The
subsequent calculation of the misfit strain energy term is identical
to that outlined in Sec. \ref{S:MBE}.

\figtwo{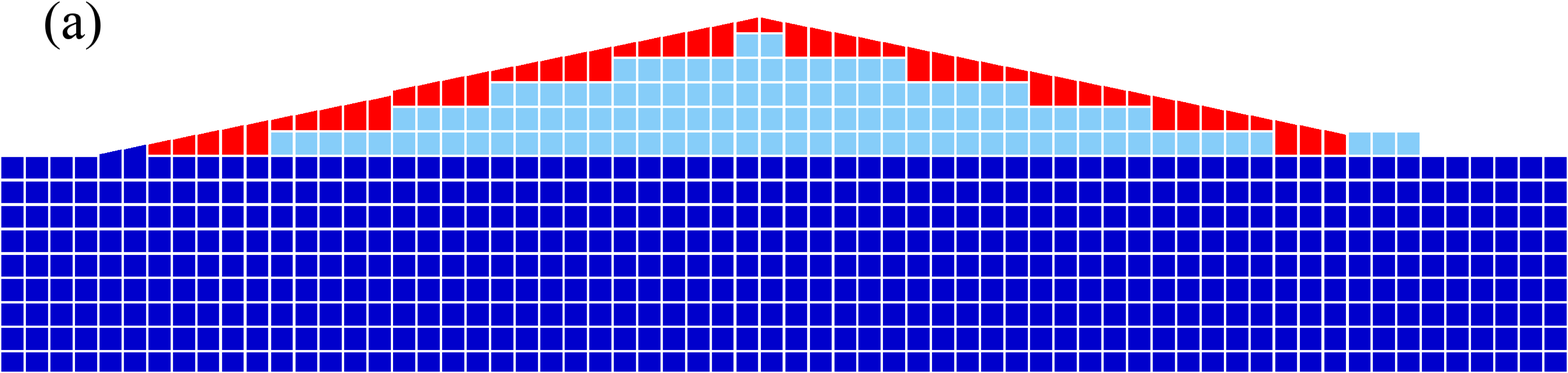}{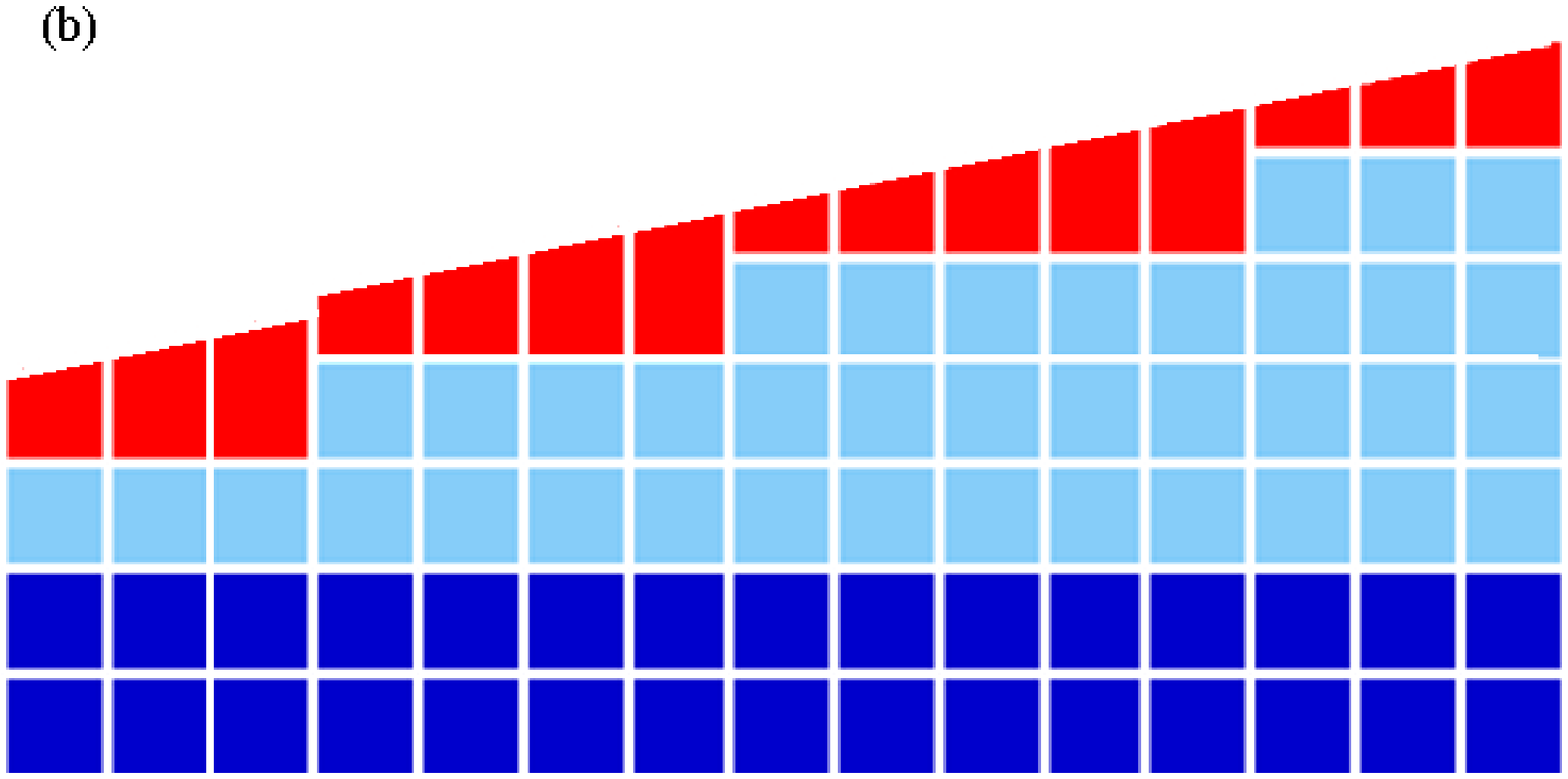}{\label{F:facet} A faceted island from a
  small scale simulation using the multi-state model (a) and a magnification
  of part of the surface containing a (105) surface step between the
  third and the fourth columns (b). Deformed film
  atoms, undeformed film atoms and all substrate atoms are shaded in
  red, light blue and dark blue respectively. In (b), the tilt variable
  $\sigma_i$ is $\frac15$ for all columns, while the extension
  variable $\kappa_i$ from left to right equals $0, \frac15, \frac25,
  - \frac15, 0, \frac15, \frac25, -\frac25, 
  - \frac15, 0, \frac15, \frac25, -\frac25, 
  - \frac15$ and 0. 
}

We first show an example of a faceted island from a small scale
simulation in Fig. \ref{F:facet}(a). Figure \ref{F:facet}(b) magnifies
part of the surface. It shows how the surface deformation smooths out
the (100) steps of the original stepped mound and turn the sidewalls
into atomically flat effective (105) facets with slopes $\pm 1/5$. An example of a
surface step on the (105) surface is also shown and will be explained
later.
In the absence of deformation, an atom is represented by a unit
square.  An integer $h_i$ denotes the surface height at column $i$. We
assume that a topmost atom in the film surface or in an exposed region
of the substrate can be deformed into a trapezoid characterized by two
new deformation state variables, namely a tilt variable $\sigma_i$ and
an extension variable $\kappa_i$.
We put 
\begin{equation}
  \sigma_i = 0,  ~ \frac15 , \mbox{~ or ~} - \frac15
\end{equation}
which gives the slope of the upper surface of the deformed atom. The
values $\sigma_i = \pm 1/5$ enable the formation of the (105)
facets in both directions. As shown in
Fig. \ref{F:facet}(b), attaining a flat (105) faceted region
further requires properly coordinated vertical stretching or
compression of the topmost atom by $\kappa_i$ which is given by  
\begin{equation}
  \kappa_i = 
    \left\{ 
        \begin{array}{ll}
            0 & \mbox{for $\sigma_i = 0$}\\
        -   \frac25, -\frac15, 0, \frac15, \mbox{~ or ~} \frac25 ~~~ & \mbox{for $\sigma_i=\pm \frac15$}
        \end{array}
    \right. 
\end{equation}
The $i$th  atomic column hence can be rectangular
or trapezoidal with the left and right edges of  
heights $h_i^a$ and $h_i^b$ given by 
\begin{eqnarray}
  h_i^a &=& h_i + \kappa_i - \frac{\sigma_i}2\\
  h_i^b &=& h_i + \kappa_i + \frac{\sigma_i}2
\end{eqnarray}
A surface step in between the $i$th and the $(i+1)$th
column has a step height $\delta_i$ defined as
\begin{equation}
  \label{delta}
  \delta_i = \mid h_{i+1}^a - h_{i}^b \mid
\end{equation}
For simplicity, we have measured step heights as projected along the lattice
axis rather than the surface normals. Note that single steps on
(100) and (105) surfaces have very different heights of 1 and 1/5
respectively in our model.

We will next explain the energy cost of the local deformation of the
surface atoms. Values of the energy parameters to be introduced are
chosen phenomenologically to provide morphologies best compared with
observations. Similar to the original lattice model \cite{Lam2002},
although we believe that our parameters are within physically
acceptable ranges, this model being in 2D is not realistic enough to
apply directly parameters from first principle studies
\cite{Montalenti2005,Lu2005,Shklyaev2005} in general.
Furthermore, we have found from numerous exploratory simulations that only a
rather limited and specific range of parameters provides reasonable
morphologies under a wide range of relevant growth conditions. The
constraints on our parameters hence may also shed light on how the
morphologies reveals certain features on the microscopic details of
the surface and this will be discussed further.

\newcommand{\Dh}{\delta_i}

The hopping rate of a topmost atom $m$ in Eq. (\ref{rate}) is generalized to  
\begin{equation}
\label{rate2}
\Gamma(m) = 
{R_0}\exp \left[ \frac{\Delta E_b(m)  + \Delta E_s(m) + E_0'}{k_{B}T}\right]
\end{equation}
where $E_0' = -\gamma-0.67$ eV. The misfit strain energy term $\Delta
E_s(m)$ is defined similarly as before and its calculation is assumed
to be completely independent of the local surface deformation. The
surface energy term $\Delta E_b(m)$ denotes the change in the bond
energy $E_b$ of the whole surface when the site is occupied versus
unoccupied. More precisely, surface energy is defined relative to that
of a flat (100) surface as
\begin{eqnarray}
\label{Eb}
  E_b  &=& \sum_i \left[ 
\eta(\sigma_i)
+  \nu(\sigma_i, \sigma_{i+1})
+\omega( \Dh , \sigma_i , \sigma_{i+1})
\right]
\end{eqnarray}
Here, $\eta(\pm 1/5)$ = 5 meV is the formation energy per site of the
(105) facet and $\eta(0)=0$ for the (100) region.  Also,
$\nu(\sigma_i,\sigma_{i+1})=0.35$ eV denotes the interface energy at
the boundary of a facet where $\sigma_i\neq\sigma_{i+1}$ and it is
zero otherwise. It dictates the energy barrier of facet nucleation.
If we choose a larger value of $\eta(\pm 1/5)$, the
(105) facet can become unstable. A negative value of $\eta(\pm 1/5)$
has been suggested \cite{Shklyaev2005} corresponding to extremely
stable (105) facets. However, this is not acceptable as island
sizes from such
simulations are then dominated by $\nu$ closely related to the edge
energy in Ref.  \cite{Shklyaev2005} but is practically independent of
the lattice misfit.

The last term in Eq. (\ref{Eb}) represents the energy of a surface
step.
On a (100) region with $\sigma_i= \sigma_{i+1}=0$, it is
defined as
\begin{equation}
  \label{omega100}
  \omega(\Dh , \sigma_i , \sigma_{i+1}) 
=           \frac{\gamma}{2}  \Dh  
\end{equation}
where the step height $\Dh$ defined in Eq. (\ref{delta}) is a integer.  This
results from simple bond counting noting that two single steps are created by
breaking one nearest neighboring bond of strength $\gamma$.  Noting
also that a bulk atom has a bond energy of $-4\gamma$,
Eq. (\ref{rate2}-\ref{omega100}) reduces exactly to Eq. (\ref{rate})
so that the (100) regions in the multi-state model behaves identically to
the basic model in Sec. \ref{S:MBE}.
Outside
of a (100) region (i.e. $\sigma_i$ or $\sigma_{i+1} \neq 0$) we put
\begin{equation}
  \label{omega105}
  \omega(\Dh , \sigma_i , \sigma_{i+1}) 
=   {\beta_{105}} \left( 
    1+\chi - \chi e^{ 1 - { 5 \Dh }} \right) +
    \frac{\gamma}{2}  \left(  \Dh   - \frac15 \right)
\end{equation}
for $\Dh\ge 1/5$ and it is zero otherwise.
This expression gives an energy $\beta_{105}$ for a single step with
height $\delta_i=1/5$ on a (105) region. It is known that incomplete
(105) facets can be practically absent at low temperature around
450$^\circ C$ \cite{Drucker2008} but are observable at 550$^\circ C$
\cite{Montalenti2004}. We reproduce this feature in our model by
taking a relatively large value of $\beta_{105}=0.3$ eV. From
Eq. (\ref{omega105}), the step energy per unit height of a multiple
step approaches $\gamma_n/2$ identical to that for a step on a (100)
facet. This also reduces the energy of an adatom on a (105) surface
which is bounded by two unit steps to a more acceptable but still very
large value of 1.3 eV. 
The parameter $\chi$ determines the energy of multiple steps of intermediate
heights. We put $\chi=0.5$ allowing a slight tendency of step bunching
\cite{Montalenti2004}.

In KMC simulation using this multi-state model, the atomic hopping events
are randomly sampled and simulated according to the rates $\Gamma(m)$
in Eq. (\ref{rate2}). We assume that the deformation state variables
$\sigma_i$ and $\kappa_i$ at every column are unchanged after an
atomic hop, i.e. the deformation state is attached to the column
rather than to the hopping atom. Deposition of an atom also increases
the column height by unity without altering the deformation state.
After every period $\tau$, the deformation state for a set of columns
will be updated.  Specifically, to facilitate program parallelization,
we adopt a sublattice updating scheme in which the deformation states
at all odd (even) lattice sites will be updated at every odd (even)
updating event.  When column $i$ is to be updated, the variables
$\sigma_i$ and $\kappa_i$ are re-sampled from the allowed set of 11
possible combinations using a heat bath algorithm based on the
relative probability $exp(-E_{b}/kT)$. We take $\tau=2/\Gamma_{ad}$
where $\Gamma_{ad}$ is the adatom hopping rate on a (100) surface
easily calculable from Eq. (\ref{rate}). This is the highest possible
rate without increasing significantly the overall execution time of
our program. Local changes in the surface reconstruction states are most
likely a fast process compared with atomic hopping. We have checked that
our deformation state updating rate is indeed sufficiently fast so that
decreasing $\tau$ gives no observable difference to our
results. Our model follows detailed balance which allows us to
confirm the reliability of our software implementation using a
Boltzmann's distribution test \cite{Lam2008}.

\section{Results}

Using both the conventional ball and spring lattice model and the
multi-state lattice model with surface deformation explained in
Secs. \ref{S:MBE} and \ref{S:extended}, we have simulated the
self-assembly of strained islands in 2D. A substrate of size 1024
$\times$ 1024 (width $\times$ depth) is used. We take a temperature
450$^\circ$ and a deposition rate 0.1 ML/s.  The conventional and the
multi-state models lead to islands with unfaceted and faceted
sidewalls respectively. For convenience, we refer to them as unfaceted
and faceted islands.

\figtwo{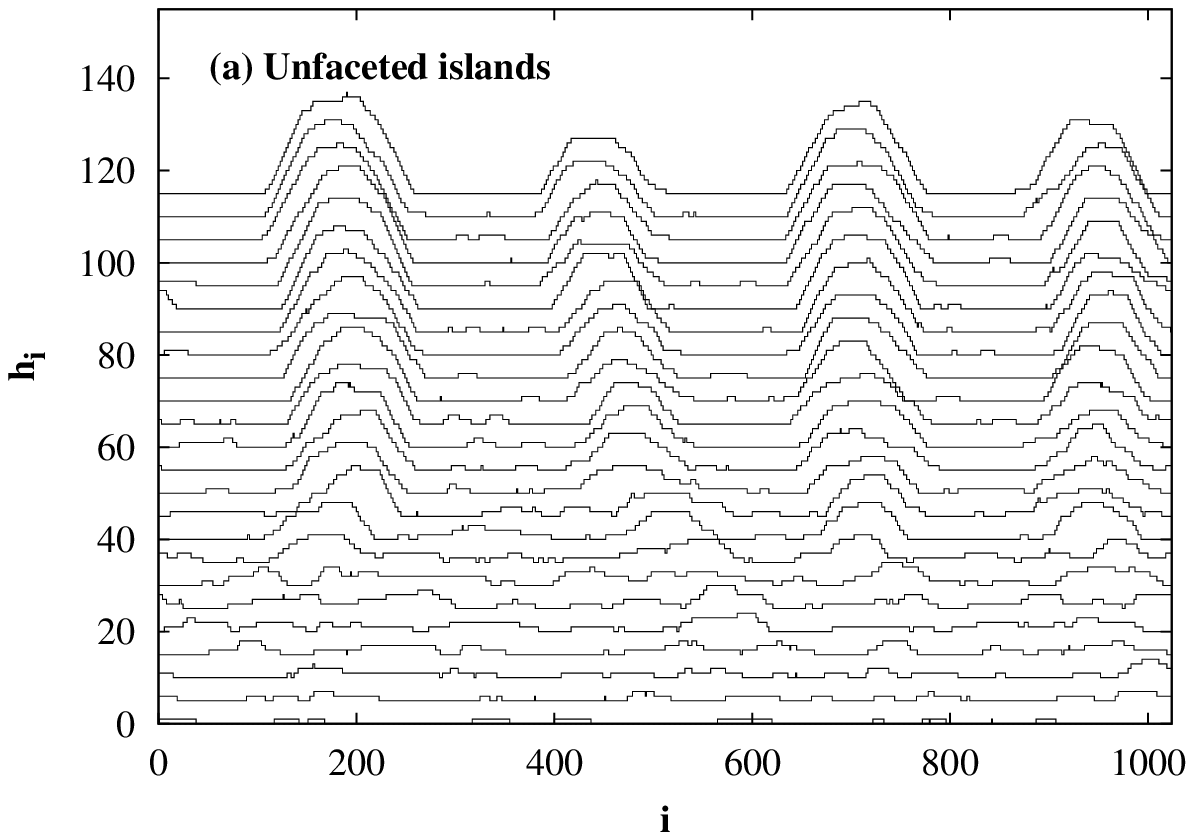}{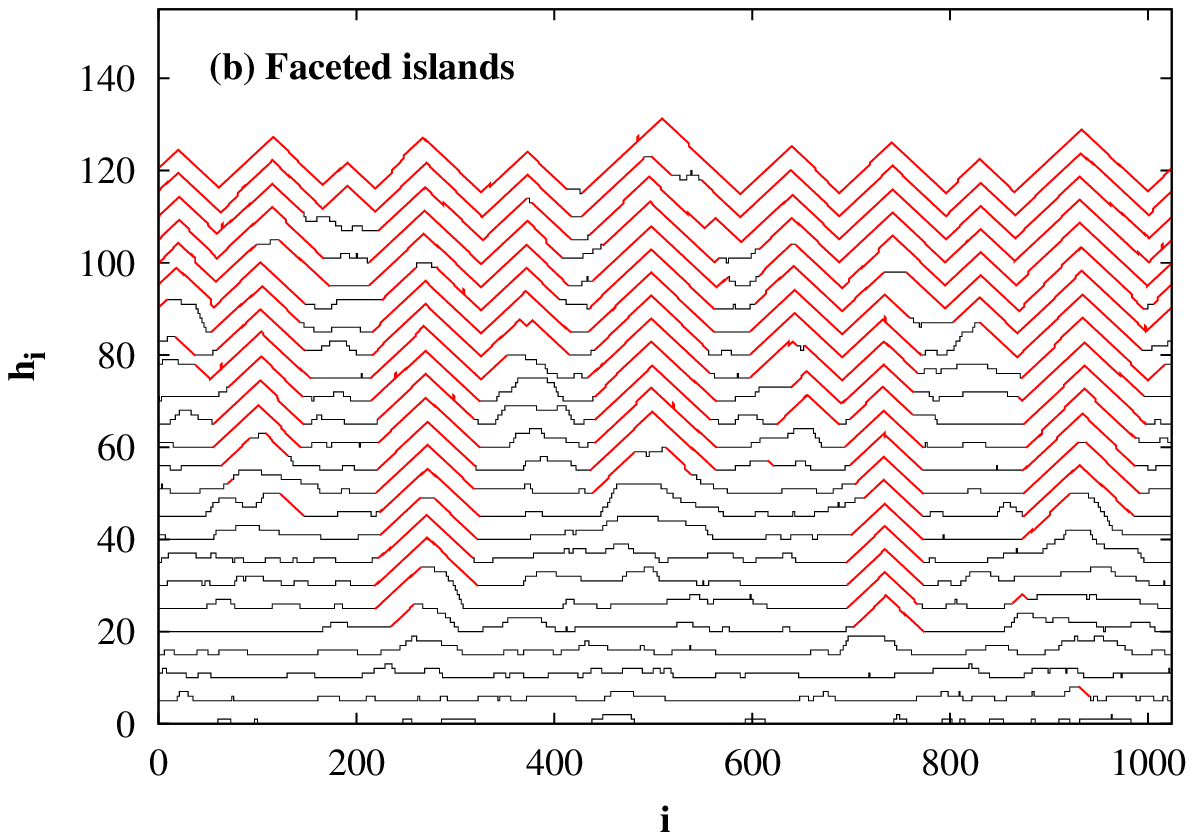}{\label{F:profile} Snapshots of
  surfaces showing the development of (a) unfaceted and (b) faceted
  islands simulated respectively using the conventional model and the
  multi-state model. (105) faceted regions
  are shaded in red. Each successive profile is displaced by $+5$
  vertically for clarity and corresponds to the deposition of a
  further 1/4 MLs up to a total of 6 MLs.}

Figure~\ref{F:profile}(a) shows the evolution of unfaceted islands
from a typical run using the conventional model during deposition of
up to 6 MLs of film material on to an initially flat
substrate. Unstable shallow stepped mounds develop at very early
stage. After depositing about 2 MLs, some stepped mounds have attained
steeper sidewalls and become more stable. At about 4 MLs, they have
generally attained the steepest possible slope of 1/2 allowed in our
model. As observed in this and other similar runs, there is a rather
well defined island nucleation period and no new island emerges after
some larger islands are established. We also observe that some
relatively mature islands eventually decay and vanish, indicating a
ripening process. The existence of a finite nucleation period followed
by ripening is consistent with previous KMC simulations \cite{Zhu2007}
as well as continuum simulations \cite{ZhangYW2003,aqua2007}. It may also have
some experimental relevance at higher temperature 
although the pyramid to dome transition and alloying between the film
and substrate atoms \cite{Rastelli2005} add further complications.

Analogous evolution of faceted islands simulated using the multi-state
model with surface deformation is shown in
Fig.~\ref{F:profile}(b). Small highly unstable (105) faceted regions
with deformed surface atoms begin to appear at a coverage of about 0.5
MLs. Relatively stable (105) faceted islands emerges at about 1
ML. These islands develops from the larger ones of the stepped
mounds. Faceted regions nucleate on either side of the mounds
independently so that half faceted asymmetric islands exist during the
course of development. Islands also often go through an truncated
pyramid stage \cite{Vailionis2000} with unfaceted tops before finally
becoming fully developed pyramids. Some faceted islands may
occasionally decay partially or even completely back to unfaceted
stepped mounds, but the larger ones are much more stable.  On the
other hand, some stepped mounds may happen to get faceted at rather
small sizes while slightly larger ones can remain unfaceted for long
periods.  Therefore, the faceting process in our current model is
strongly affected by both the energetics and the kinetics.

At this low growth temperature of $450^\circ$, surface steps on a
(105) facet is rare as explained in Sec. \ref{S:extended}. Further
growth of faceted islands by step flow is hence kinetic limited
\cite{Kastner1999,Drucker2008}. It can be observed from
Fig.~\ref{F:profile}(b) that island growth rates drop dramatically
once becoming faceted. Their sizes occasionally jump up rapidly only
when parts of the sidewalls become temporarily unfaceted due to
thermal excitation. Since developed islands are poor absorber of newly
deposited atoms, new islands continue to nucleate until the substrate
is crowded with islands. Kinetic limited growth and continuous island
nucleation have not been reported previously in KMC or continuum
simulations in our knowledge. More importantly, deposition experiments
at 550$^\circ$C do indicate slower growth of matured islands and a
continuous island nucleation growth mode \cite{Rastelli2003}. Deposition at lower
temperature however leads to huts \cite{Kastner1999,Drucker2008} which
may share some related characteristics but are more complicated.

\fig{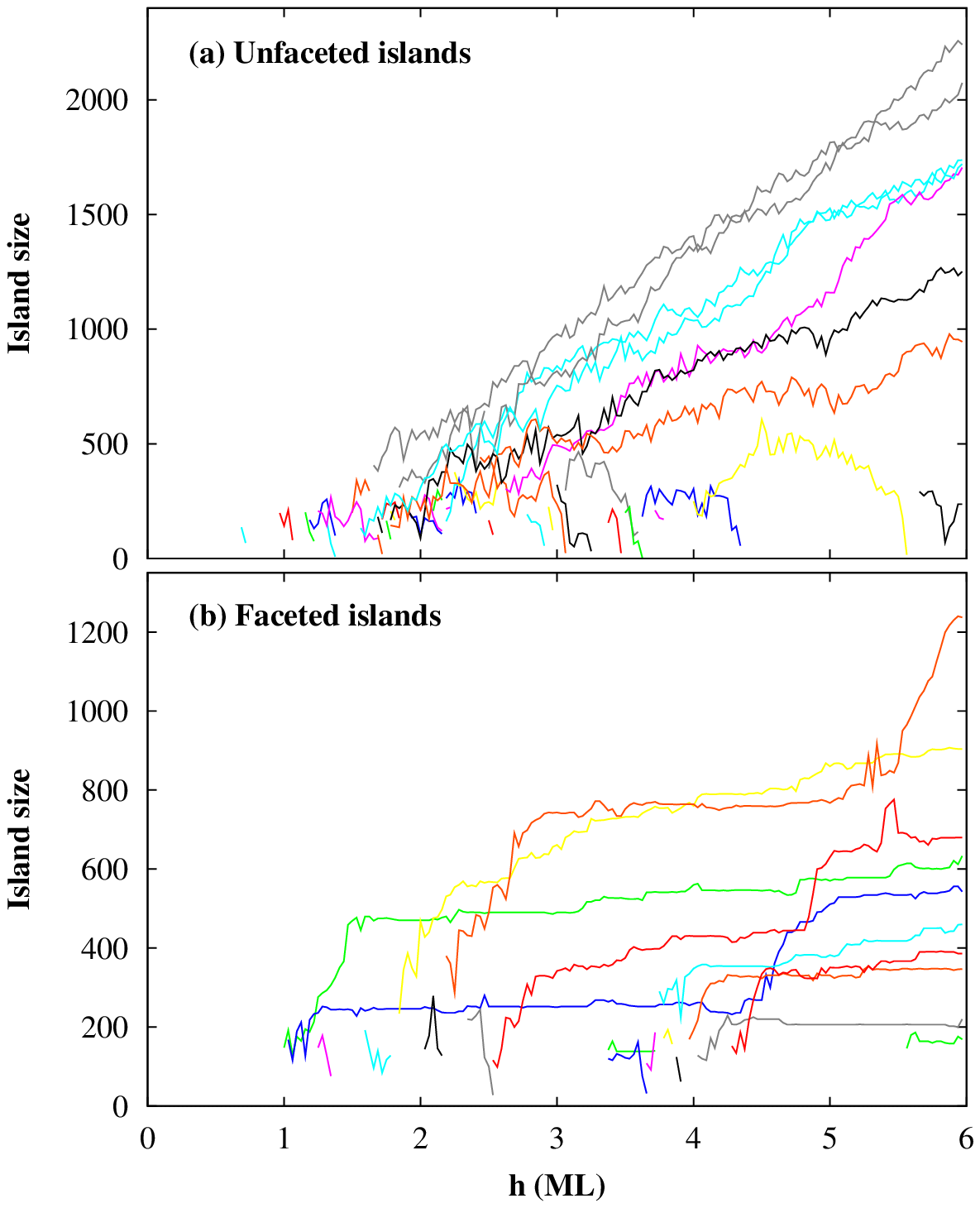}{\label{F:size} Plot of island size against nominal film
  thickness $h$ for unfaceted (a) and faceted (b) islands }

For more quantitative analysis, we define an island as one in which
each of the constituent columns must be at least 4 atoms tall. All
islands can then be automatically identified. Figure~\ref{F:size}
traces the size evolution against the nominal film thickness $h$ of
every island in Fig.~\ref{F:profile} once they have attained a size of at
least 150 atoms. Islands from another
similar run are also included in Fig.~\ref{F:size}(b) to provide additional examples. From
Fig.~\ref{F:size}(a), unfaceted islands beyond a certain size in
general grow steadily with its own characteristic rates which are
expected to depend mainly on the sizes of their adatom capture
zones. Small islands decay and vanish. In contrast, from
Fig.~\ref{F:size}(b), there is in general an initial period of rapid
island growth followed by much slower growth after faceting. Once
faceted, their sizes remain nearly constant except at occasional jumps
associated with temporary partial decay of the facets as described
above.

\fig{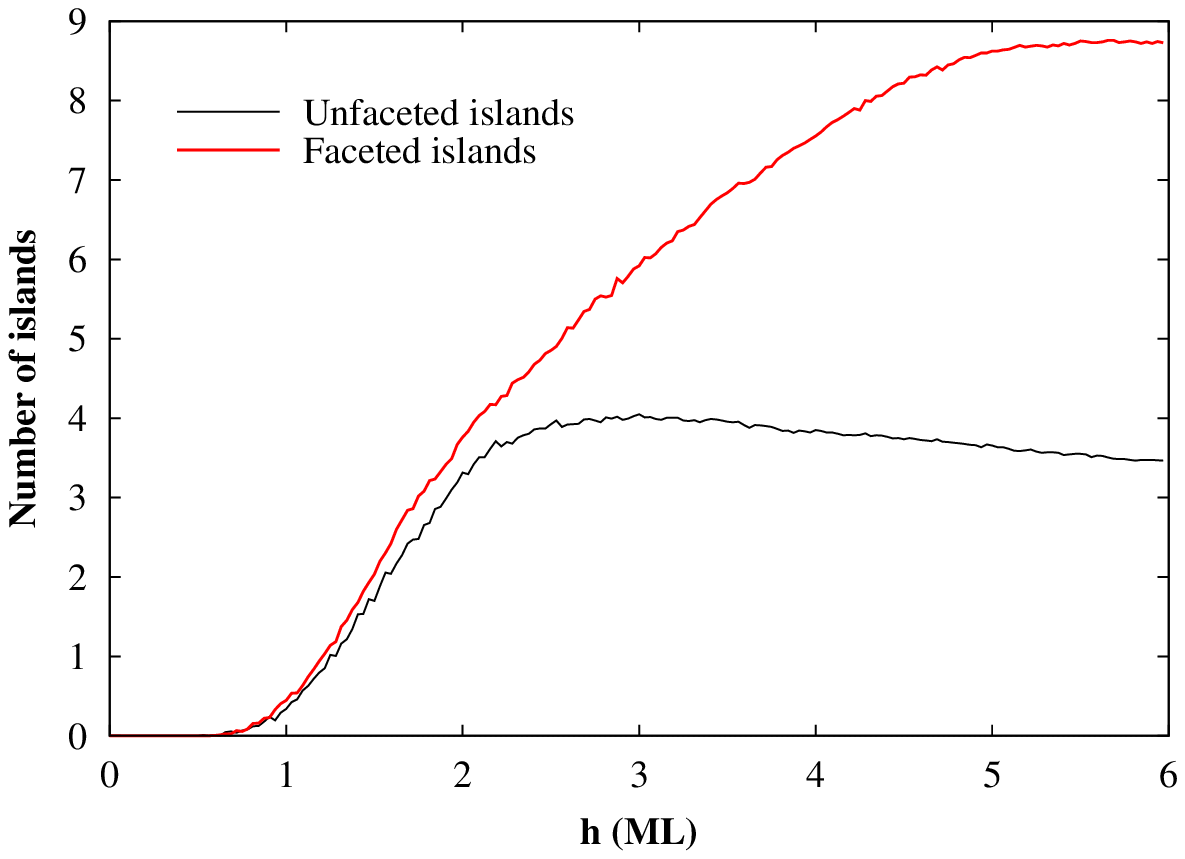}{\label{F:number} Plot of the average number of
  islands on a substrate of 1024 atoms wide against nominal film
  thickness $h$ }

To obtain more statistics, we have repeated each simulation 200
times. Figure~\ref{F:number} plots the average number of islands of
size 150 or larger on the 1024 atoms wide substrate used. Smaller
islands are excluded because they are highly unstable. For unfaceted islands, their
number first increases indicating a period of active
nucleation at coverage from about 1 to 2.5 MLs. It 
then declines but at a very slowly rate indicating rather inefficient
coarsening during growth.
In contrast, faceted islands steadily increase in number for
coverage up to about 5 MLs due to continuous nucleation. 
Beyond 5 MLs, the substrate is crowded with islands and the
number of islands saturates.

\figtwo{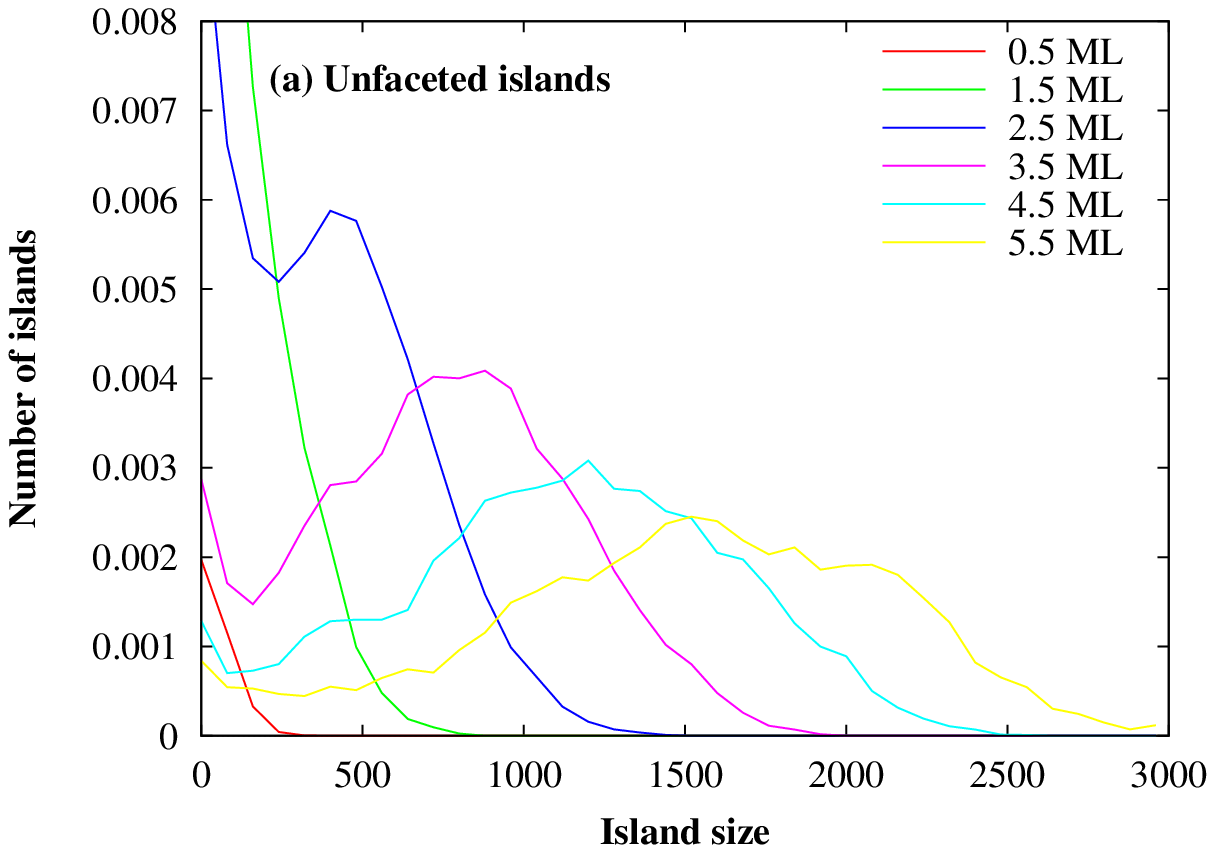}{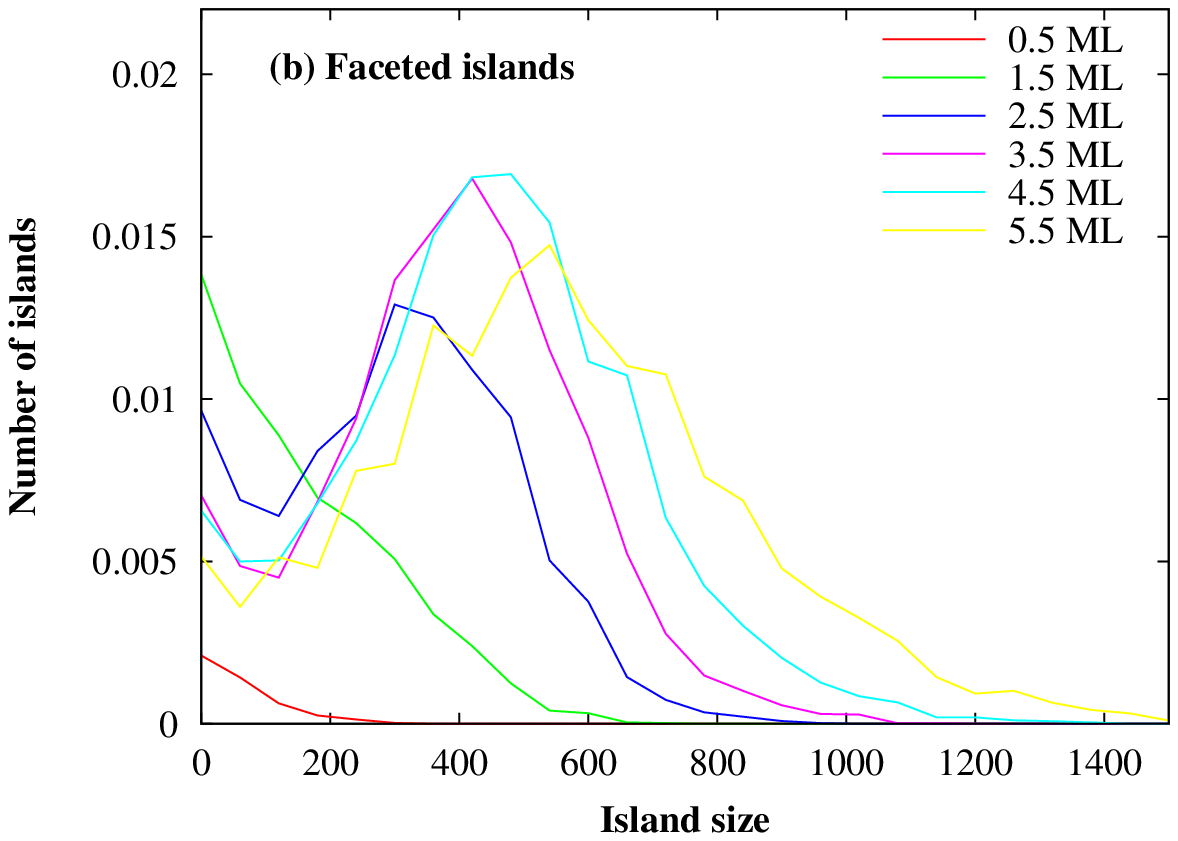}
{\label{F:histo} Size histograms for unfaceted (a) and
  faceted (b) islands}

Finally, we histogram the island sizes from all the independent runs.
Fig.~\ref{F:histo} plots the average number of islands on the
substrate against island size. For both models, a peak island size
emerges for $h\agt 2.5$ MLs.  For unfaceted islands, the histogram
broadens significantly upon growth due to a wide distribution of
growth rates. In contrast for faceted islands, it broadens much more
slowly due to the highly kinetic limited growth mode. Nevertheless,
the faceted islands do not possess narrower size distribution relative
to the average size. This is because a significant size distribution
already exists when the islands become faceted as can be observed in
Fig. \ref{F:profile}(b). The continuous nucleation of new islands also
broadens the distribution as the older islands are larger on average.
Another difference between the models is that the peak of the histogram
decays monotonically upon deposition for unfaceted islands while it
increases for $2.5 \le h\le 4.5$ due to the continuous
nucleation of islands.

\section{Discussions}

We have generalized a lattice model for strained films to allow for a
range of local deformation states of surface atoms representing
effective surface reconstructions. This deformations are assumed to be
independent of the misfit induced strains for
simplicity. Using this multi-state lattice model, we have
performed kinetic Monte Carlo simulations in 2D and observed the
formation of (105) faceted pyramid islands. 
The model enables us to simulate faceted island formation in the
kinetic limited regime. In this regime, island growth slows down
dramatically and becomes intermittent after faceting. The slower
growth of the more established islands also leads to a
continuous nucleation of islands until the substrate is fully
occupied.  The width of the island distribution is dominated both by
fluctuations in the initial size at the start of faceting as well
as the diversity in their ages. Stepped mounds from the
conventional model exhibit a simple nucleation period followed by
slow ripening.

Additional studies on the growth and annealing of faceted islands under
other growth conditions will be report elsewhere. It is also
interesting to further generalize the model to consider two facet types
so as to study the pyramid to dome transition.
Generalization to 3D is conceptually straightforward but is
challenging in practice because of the heavy computational load expected. 

This work was supported by HK RGC, Grant No. PolyU-5009/06P. 

~\\~\\~\\~\\~\\~\\~\\~\\~\\~

\bibliography{island}

\end{document}